# Photoconversion of shallow nitrogen-vacancy centers in flat and nanostructured diamond under high-power laser irradiation

Federico Gorrini[1,2], Carla Dorigoni[3], Domingo Olivares-Postigo[2,3,6], Rakshyakar Giri[3], Pietro Aprà[4,5], Federico Picollo[4,5] and Angelo Bifone[1,2,6]

[1]Istituto Italiano di Tecnologia, Center for Sustainable Future Technologies, via Livorno 60, 10144 Torino, Italy
[2]University of Torino, Molecular Biology Center, via Nizza 52, 10126, Torino, Italy
[3]Istituto Italiano di Tecnologia, Center for Neuroscience and Cognitive System, corso Bettini 31, 38068, Rovereto (Tn), Italy
[4]University of Torino, Department of Physics and "NIS inter-departmental centre", Via Pietro Giuria, 1, 10125 Torino, Italy
[5]National Institute of Nuclear Physics, Section of Torino, Torino 10125, Italy
[6]University of Torino, Department of Molecular Biotechnology and Health Sciences, via Nizza 52, 10126, Torino, Italy

**(Abstract)** Shallow, negatively-charged Nitrogen-Vacancy centers ($NV^-$) in diamond have been proposed for high-sensitivity magnetometry and spin-polarization transfer applications. However, surface effects tend to favor and stabilize the less useful neutral form, the $NV^0$ centers. Here, we report the effects of green laser irradiation on ensembles of nanometer-shallow NV centers in flat and nanostructured diamond surfaces as a function of laser power in a range not previously explored (up to 150 mW/$\mu m^2$). Raman spectroscopy, optically-detected-magnetic-resonance (ODMR) and charge-photoconversion fluorescence detection are applied to characterize the properties and dynamics of $NV^-$ and $NV^0$ centers. We demonstrate that high laser power strongly promotes photoconversion of $NV^0$ to $NV^-$ centers. Surprisingly, the excess $NV^-$ population is stable over a time scale of one hundred milliseconds after switching off the laser, resulting in long-lived enrichment of shallow $NV^-$. The beneficial effect of photoconversion is less marked in nanostructured samples. Our results are important to inform the design of samples and experimental procedures for applications relying on ensembles of shallow $NV^-$ centers in diamond.

## Introduction

Negatively-charged nitrogen-vacancy ($NV^-$) centers are solid-state defects in the diamond lattice whose properties have been exploited to detect temperature gradients[1,2], magnetic[3,4] and electric fields[5,6] at the nanoscale, as well as interactions with magnetic molecules and nanoparticles[7–10]. Due to their biocompatibility, $NV^-$—enriched fluorescent nanodiamonds represent promising sensors to investigate the cellular microenvironment in living tissues, and their use in high-sensitivity bioassays has been proposed[11–13]. Furthermore, NV centers can be used in dynamic nuclear polarization (DNP) protocols where the polarization of the $NV^-$s is transferred to $^{13}C$ nuclei, leading to hyperpolarization of the $^{13}$C nuclei in the diamond lattice[14–16]. Substantial efforts are ongoing to promote polarization transfer from shallow NV centers to molecules absorbed at the diamond surface, thus enabling hyperpolarization of high-sensitivity tracers for biomedical Magnetic Resonance Imaging.

For all these applications, proximity of NV$^-$ centers to the diamond surface, where NVs can effectively interact with spins outside the diamond lattice, is of paramount importance, as the coupling strength between magnetic dipoles decreases with increasing distance. To this end, specially engineered layers of shallow NV$^-$s have been proposed[17,18], as well as nanodiamonds[19].

Unfortunately, surface states and defects at and close to the diamond surface can affect charge stability of NV centers, reducing the availability of magnetically active NV$^-$ centers in favor of the neutral form (NV$^0$ centers), which do not present the same detection features. The relative stability and interconversion between the neutral and negatively charged states of the NV centers have been the object of investigation in several studies[20–24], and various attempts have been made to increase the stability of NV$^-$ centers, for instance by surface termination[25], by doping of the diamond lattice[26] or by application of an electric field[27].

Here we investigate the effects of laser power and surface structure on charge stability and attainable spin polarization of shallow NV$^-$ centers in high-purity diamond. Specifically, we aim to establish experimental conditions that maximize availability of magnetically active NV$^-$ at the diamond surface. To this end, we apply fluorescence and Raman spectroscopy, as well as Optically Detected Magnetic Resonance (ODMR), in electronic-grade diamond samples implanted with nitrogen ions at different depths, and with flat or nanostructured surfaces.

Importantly, we study the effects of these factors in a regime of high-power laser irradiation not previously explored, and we assess the time dependence of the laser-induced NV$^-$/NV$^0$ distribution during and after irradiation.

**Methods**

*Samples fabrication*

We used four electronic-grade, single-crystal diamond plates, with initial nitrogen concentration lower than 5 ppb (Element Six Technologies Ltd). Samples were 2.0x2.0 mm, 0.5 mm thick, with {100} face orientation.

Two of these samples were nanostructured to increase the surface-to-volume ratio. Samples nanostructuring was carried out by means of Inductively Coupled Plasma - Reactive Ion Etching (ICP-RIE) with oxygen gas, using a non-continuous gold layer as mask[28]. Gold was deposited by DC sputtering (KS500 confocal sputtering system - Kenosistec Srl), using Ar gas (Fig.1a). Etching was performed by means of ICP-RIE (SI500 - Sentech Instruments GmbH), using 700W ICP power, 50 W plate power and 40 sccm O$_2$, at 1 Pa pressure, for 60 sec (Fig.1b). After the ICP-RIE process, gold was removed by means of commercial gold etchant (Sigma Aldrich). Scanning Electron Microscopy (SEM) shows a tight arrangement of nanostructures ("nanopillars") (Fig.1e) over the diamond surface, with a height of ≈150 nm. SEM images were acquired before gold removal.

Samples were implanted with $^{15}$N (INNOVION Corp - San Jose, CA) at a fluence of $10^{13}$ cm$^{-2}$, at 7° angle from normal to avoid ion-channeling[29–31] (Fig.1c), in two different energy conditions: one flat sample and one nanostructured sample were implanted at 10 keV; the other flat and nanostructured ones were implanted at 20 keV (see Table 1). Ion average ranges in the flat samples calculated by SRIM (http://www.srim.org, SRIM-2013.00) are respectively ≈15 nm, with straggling of 5 nm, and ≈27 nm with straggling of 8.5 nm[32].

We then annealed the four samples at the temperature of 850 °C for 2 hours in high vacuum conditions ($10^{-6}$ mbar) (Fig.1d). The set-up employed consists of a vacuum chamber equipped with a dry pumping system, to avoid hydrocarbon contamination, and a resistive heater. The heating element is a tantalum box, 5×5×5 mm$^3$ in size, that keeps the sample in thermal equilibrium with the surrounding black-body radiation. The temperature was externally monitored through an optical window using a pyrometer, pointed to the tantalum box. The heating and cooling rates were of 10 °C min$^{-1}$ to avoid thermal stresses in the diamond structures. At this temperature, vacancies and interstitial nitrogen atoms become mobile, but substitutional nitrogen atoms are fixed in their position[33]. We expect deeper NV centers for the 20 keV than for 10 keV implantation energies on average. In the nanostructured samples a more inhomogeneous distribution profile is expected, as the implantation occurs on an irregular profile.

The fabrication process for the nanostructured samples is summarized in Figure 1a-d. For flat samples implantation and annealing procedures were the same as in the nanostructured samples, but performed on the flat surface.

| Sample name | Surface type | Implantation energy (keV) |
|---|---|---|
| F1 | flat | 20 |
| F2 | flat | 10 |
| N1 | nanostructured | 20 |
| N2 | nanostructured | 10 |

**Table 1 – Type of surface and implantation energy of $^{15}$N$^+$ for the four electronic-grade samples used in this work.**

*Experimental setup*

Full fluorescence spectra of the NVs were taken with a confocal microRaman setup (LabRam Aramis, Jobin-Yvon Horiba), equipped with a 532 nm DPSS laser and an air-cooled multichannel CCD detector. Laser power was attenuated by neutral filters with 0, 1 or 2 optical densities (OD).

Time-resolved measurements were performed with a home-built fluorescence microscope equipped with a green 532 nm laser (5 W, Verdi, Coherent) and a 0.25 NA objective (Plan N, Olympus). The focal spot size is estimated between 6 and 13 $\mu m^2$. We used the same 532 nm laser to initialize and readout the state of the system. Fluorescence was collected by the objective and sent to a photon-counter module (Excelitas SPCM-AQRH-14-FC). Laser power at the sample surface was attenuated with a combination of absorptive filters, and ranged from a maximum value of 1.5 W to 1 mW, depending on the experiment's purpose.

We collected fluorescence through two different spectral windows, selected by combinations of filters: 550 nm to 600 nm ("550-600nm") and >750 nm ("750nm+"). The 550-600nm window is centered around the zero-phonon-line (ZPL) of the $NV^0$ centers, at 575 nm, and collects exclusively the $NV^0$ signal. Conversely, the 750nm+ window is more selective for the $NV^-$ centers, even if a tail of the $NV^0$ fluorescence can still leak through (see Appendix I).

We used a high-power acousto-optic modulator (ISOMET 523C-6) for pulsed-laser sequences. Typically, a variable preparation pulse was applied to initialize the system, followed by a read-out pulse. A read-out pulse of 5 µs provided good sensitivity while minimally perturbing the distribution of NV charge states even at the highest laser power used (see Appendix II).

The microwave lines comprised a microwaves generator (Keysight N5171B), an amplifier (ZHL-16W-43-S+) and a millimeter-sized gold-coated copper loop placed below the sample. A custom three-axis Helmholtz coil (Micro Magnetics, Inc) provided a static magnetic field along any desired direction. In some experiments, a strong magnetic field of 750 G was applied along the z-axis, perpendicularly to the top diamond face, to quench the spin dynamics and enable selective measurement of charge dynamics[34]. The z-axis was set parallel to the [100] crystallographic direction, so that the applied magnetic field had the same magnitude along the four possible orientations of the NV centers axes in the diamond lattice.

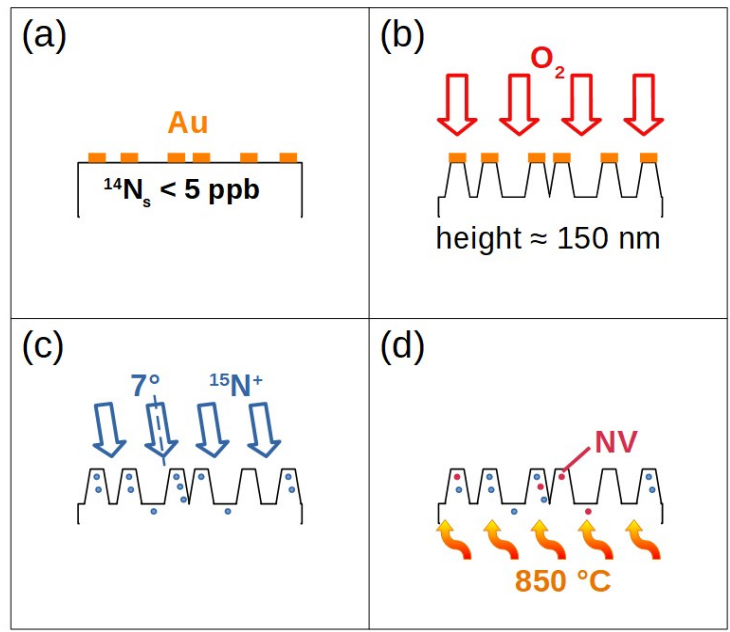
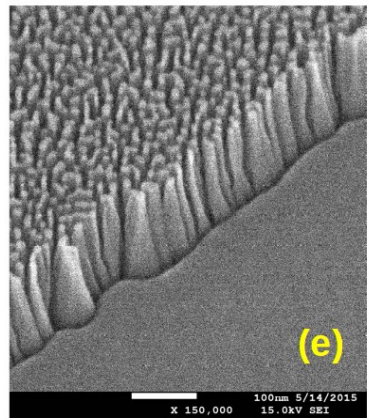

*Figure 1 – **Nanofabrication process and synthesis of shallow NV centers in nanostructured samples.** Electronic grade diamond with very low concentration of nitrogen, less than 5 ppb, was masked with gold by sputtering (a) and then etched by using oxygen reactive ion etching (b) to fabricate nanostructures, onto which nitrogen was implanted after gold removal (c). Samples were then annealed (d). SEM images show ≈150 nm high nanopillar-like structures (e). The SEM image was taken between steps (b) and (c), with still some conductive gold on the tips, to improve image quality. Flat samples were implanted with $^{15}N$ ions and then annealed, without application of any etching procedure.*

## Results

Figure 2 shows the effects of laser power on the fluorescence (FL) spectra from different samples. Schematics of the energy level structures of $NV^0$ and $NV^-$ centers are reported in the panel Fig.2a, to facilitate interpretation of the experimental results. In short, NVs are characterized by a ground state (a spin triplet $^3A_2$ for the NV⁻, and $^2E$ for NV⁰) and excited states ($^3E$ for NV⁻ and $^2A_1$ for NV⁰) positioned within the diamond band gap[35], each accompanied by a phonon band. Additionally, the NV⁻ has two intermediate states ($^1A_1$ and $^1E$) coupled to the ground and the exited states. The existence of this coupling enables a non-radiative transition that accumulates population in the $m_s=0$ level of the ground state under constant laser irradiation (blue arrows). A green laser (532 nm) is used to excite both the NV⁻ and the NV⁰ centers (green arrows). Laser irradiation can also induce a charge-state conversion, either by promoting an electron from the NV⁻ excited state into the conduction band (NV⁻→NV⁰ conversion) or by exciting an electron from the valence band to the NV⁰ ground state (NV⁰→ NV⁻ conversion)[36]. These two routes of photo-conversion are represented by the green dotted arrows. Experimental observations suggest that, upon switching off the laser, the system goes back to equilibrium through charge conversion in the dark, a process attributed to electron tunneling (black dotted arrows). Indeed, tunneling couples the NV centers to proximal electron donor or acceptor states, such as substitutional nitrogen atoms, vacancy complexes and surface states[22,24,37].

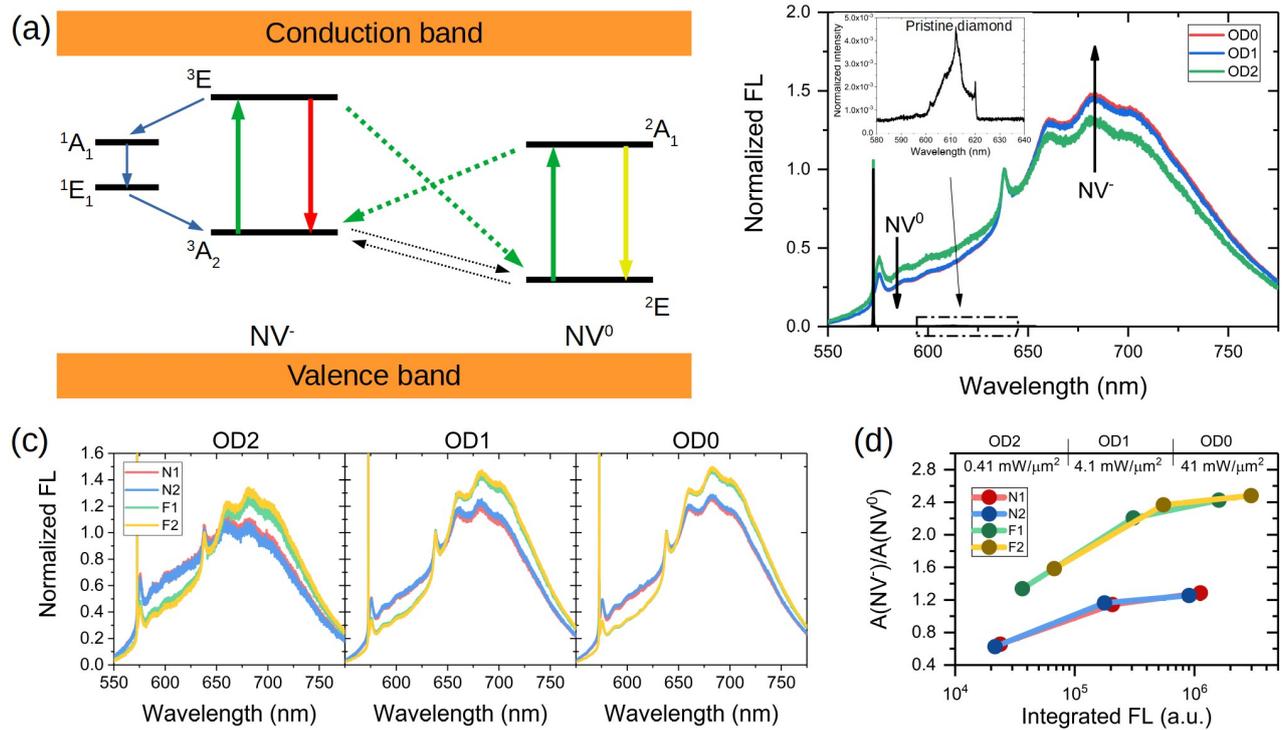

*Figure 2 – Fluorescence properties of NV centers. Electronic structure of NV⁻ and NV⁰ centers is represented in (a). Green arrows indicate optical excitation and photoconversion (continuous and dotted arrows, respectively). Red and yellow arrows denote radiative decay from NV⁻ and NV⁰ centers. Blue arrows indicate the non-radiative pathway which leads to polarization of the $m_s=0$ NV⁻ ground state. Black arrows represent tunneling transitions in the dark between the two charged states. (b) Both NV⁻ and NV⁰ fluorescence spectra were detected in implanted samples; no signal was detectable in non-implanted, non-annealed samples (inset). Notably, emission from NV⁻s*

*increases with laser power, at the expenses of the $NV^0$s. (c): the FL of nanostructured samples is lower compared to flat samples at all laser power levels (indicated by the filter optical density OD), with a larger component of $NV^0$. The ratio of $NV^-$ and $NV^0$ FL intensities is plotted in (d), as a function of total integrated fluorescence and laser intensity. The $NV^-/NV^0$ ratio increases with laser power in all samples, and is systematically larger in flat samples.*

Figure 2b shows fluorescence spectra from one sample (F2) at different laser powers. NV charge states are characterized by different fluorescence spectra, with zero-phonon lines (ZPL) at 575 nm and 638 nm for the $NV^0$ and $NV^-$, respectively, and a phonon sideband peaked around 620 nm for the $NV^0$ and 700 nm for the $NV^-$, and extending up to ≈800 nm in both cases. Spectral features from both $NV^-$ and $NV^0$ centers are apparent, demonstrating the co-occurrence of the two charge states in this sample. The NV centers originate exclusively from implanted nitrogen (pristine diamond samples before implantation and annealing show no detectable fluorescence, but only the first and second order Raman peaks, shown in the inset). As the laser power intensity increases by two orders of magnitude (from 0.41 mW/µm² for filter optical density OD=2 to 41 mW/µm² at OD=0), the relative intensity of the $NV^-$ band increases with respect to that of the $NV^0$ band. This indicates some degree of $NV^0 \rightarrow NV^-$ photoconversion at higher laser powers under continuous excitation (curves are normalized to 1 at 638 nm). The spectra in Figure 2c reflect the relative abundances of $NV^0$ and $NV^-$ centers in the nanostructured and flat samples for different laser powers. For the two nanostructured samples, overall FL is lower (not apparent in the figure, where spectra are normalized for comparison), and the FL spectra show a higher component in the $NV^0$ region (blue and red curves compared to yellow and green curves). In all samples, an increase in the $NV^-$ component and a decrease in $NV^0$ is observed with increasing laser power. We summarize these observations in Fig 2d, where the x-axis represents the integrated total FL, comprising the signals from both NV types, while the y-axis indicates the ratio of the areas under the $NV^-$ and $NV^0$ curves. For all samples, the $NV^-/NV^0$ ratio increases with laser power, suggesting stronger $NV^0 \rightarrow NV^-$ photoconversion at higher laser powers. Nanostructured samples (N1 and N2) contain fewer NV centers (reflected by an overall lower FL, integrated over the entire spectrum), and a higher fraction of $NV^0$ compared to flat samples.

In order to explore the dynamics of laser-induced charge switching and subsequent return to equilibrium, we applied two pulsed experimental schemes (Fig.3). In these experiments, a strong magnetic field (750 G) was applied along the [100] direction to suppress the effects of spin polarization of FL and selectively investigate charge dynamics. In fact, a sufficiently strong magnetic field (>600 G) quenches the polarization of the $m_s$=0 ground state by spin-state mixing, and leads to a reduction of the spin-related component in the FL[34,38,39].

Figure 3a and 3c show the two pulse sequences adopted to assess charge dynamics during laser irradiation and in the following dark time. In the first one (Fig.3a) a laser pulse (532 nm) of variable duration $\tau_L$ (from 10 µs to 100 ms) was focused on the diamond surface to study charge photoconversion during irradiation. This pulse induces photoionization, with an increase in $NV^-$ and a decrease in the $NV^0$ signal (blue and orange curves, respectively). FL was recorded by the subsequent short detection pulse of 5 µs. At the end of the sequence, a time interval of 10 ms, during which the laser was off, allowed for charge relaxation before the experiment was repeated for signal averaging. Data points acquired after discrete initialization pulses, for different values of laser power, are shown in Fig.3b (for sample F2). For simplicity, only FL signals acquired in the range 550-600nm, reflecting predominantly $NV^0$ fluorescence, are reported (the behavior of $NV^-$ is symmetrical). The curves reported in this figure show that the degree of photoconversion increases steeply with lasers power, ranging between 50 mW to 1400 mW. Independently of laser power, for pulses longer than 10 ms, the fluorescence level lies within 5% of its equilibrium value, and photoconversion approaches saturation in all samples. Therefore, in the following experiments we use an initialization pulse of at least 10 ms to maximize the number of $NV^-$.

The second pulse sequence of Fig.3c was used to probe the charge dynamics in the dark. The sequence consists of a laser pulse of duration $\tau_L$ followed by a variable dark time $\tau_D$ (from 1 µs to 10 ms) and a 5 µs readout pulse. In Fig.3d, the FL of the $NV^0$ centers measured through the 550-600nm spectral window are reported as a function of $\tau_D$. As discussed above, the first initialization pulse reduces the number of $NV^0$ by photoconversion to $NV^-$. During the dark time the system goes back to equilibrium, and a steady increase in the number of $NV^0$ centers is reflected by increased fluorescence in the spectral region 550-600nm registered by the readout pulse. A plausible explanation of this recharging-in-the-dark process, previously reported and discussed[22,24], contemplates electrons tunneling from the negative $NV^-$ to surface acceptors or vacancies until equilibrium is reached. To exemplify this phenomenon, Fig.3d shows the curves obtained with a lasers power of ≈625 mW and initialization times from 10 µs to 100 ms for sample F2. Coherently with the results of Fig.3b, the initial value of fluorescence decreases with longer initialization pulses, meaning that an increasing fraction of $NV^0$ is converted into $NV^-$ upon laser irradiation. In all cases, fluorescence from $NV^0$ increases in the dark with longer $\tau_D$. We also notice that the curves acquired after a $\tau_L$ of 10 ms, 30 ms and 100 ms are almost identical, consistent with the observation that beyond 10 ms the photoconversion process saturates.

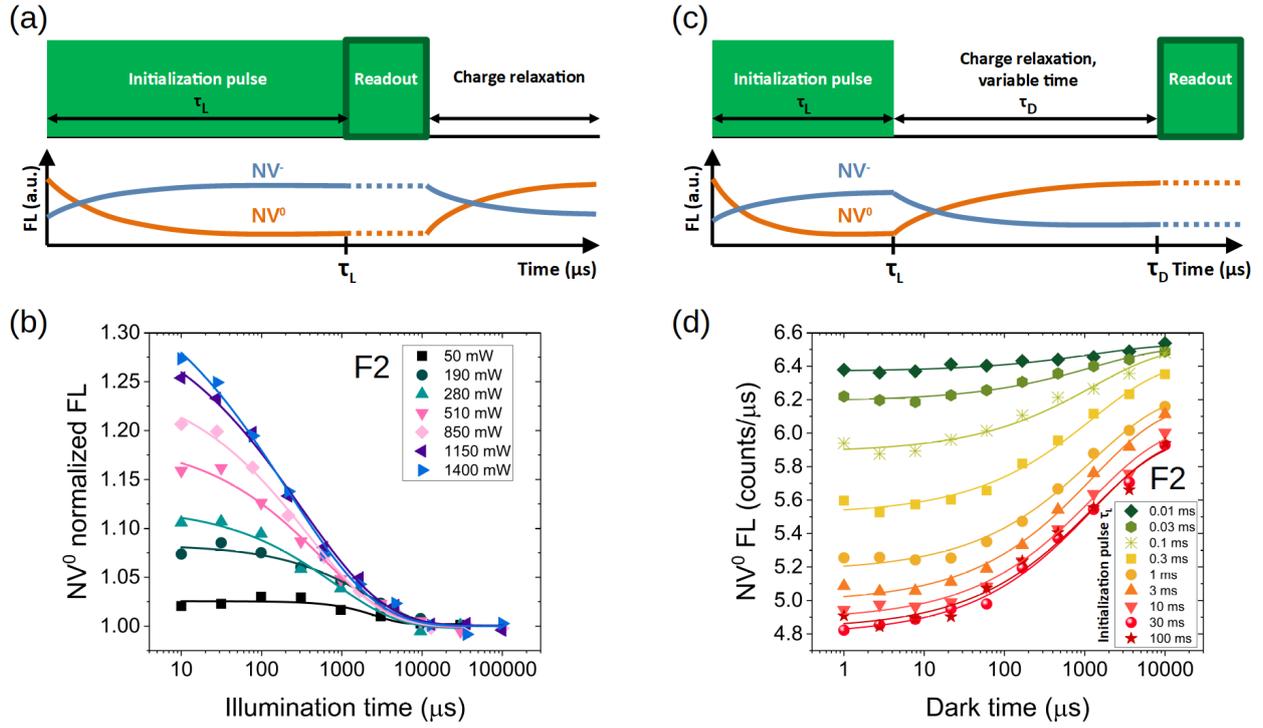

*Figure 3 – Detection of charge dynamics. (a): Pulse sequence adopted to investigate charge dynamics under laser pumping, for variable irradiation pulse length $\tau_L$. Laser induced $NV^0 \rightarrow NV^-$ photoconversion results in a decrease of the $NV^0$ signal and an increase of the $NV^-$. As an example, panel (b) shows the decrease in $NV^0$ FL for sample F2, as measured in the 550-600nm window, for different laser powers. (c) Pulse sequence used to detect the charge recovery in the dark after a preparation pulse. Recovery of $NV^0$ FL for sample F2 at laser power 625 mW is shown in panel (d). The various curves in panel (d) correspond to different durations of the initialization pulse.*

The curves of Fig.3b can be fitted with a stretched exponential law of the type[8]

$$I(t) = I_{eq}\left[1 + C\,e^{-(t/T_r)^n}\right] \qquad \text{(Eq. 1)}$$

where $T_r$ represents a characteristic timescale for the photoconversion process under the laser pulse, n is the stretching factor, $I_{eq}$ is the FL value at equilibrium and C is a positive parameter describing the fluorescence drop from the initial value to equilibrium. Curves were normalized by $I_{eq}$ for easier comparison. Charge recovery curves, as in Fig.3d, can also be fitted with the function of Eq.1, even though, in this case, the pre-exponential parameter is negative (the initial value is lower than the equilibrium value). It should be noted that the mechanisms underlying charge conversion in the dark are different from those that induce photoconversion, so the two characteristic times $T_r$ are not related.

The photoconversion data for all samples are reported in Fig.4a, where we plot $(1+C)^{-1}$ (Eq.1), corresponding to the ratio between the FL equilibrium and initial values, for different laser powers.

The largest drop in $NV^0$s due to photoconversion was observed in the flat samples (down to 74% for F2 and 81% for F1) at the largest available laser power, while more limited photoconversion was observed in the nanostructured samples. Note that the large error bars on the x-axis are caused by a slight deterioration in the laser beam's transmittance at high laser power, an effect that accounts to a 5-10% of losses during prolonged irradiation. In Fig.4b we show the rate of photoconversion (i.e. the inverse of the photoconversion time $T_r$) for the flat samples (in the case of nanostructured samples the uncertainties on extracted values are too large to draw conclusions on their laser-power dependence). $(T_r)^{-1}$ is seen to increase with laser power according to a power law with exponents of 0.95±0.07 and 1±0.07 for samples F1 and F2, respectively (dashed lines). The nearly linear dependence on the laser power is suggestive of a single-photon-mediated photoconversion mechanism.

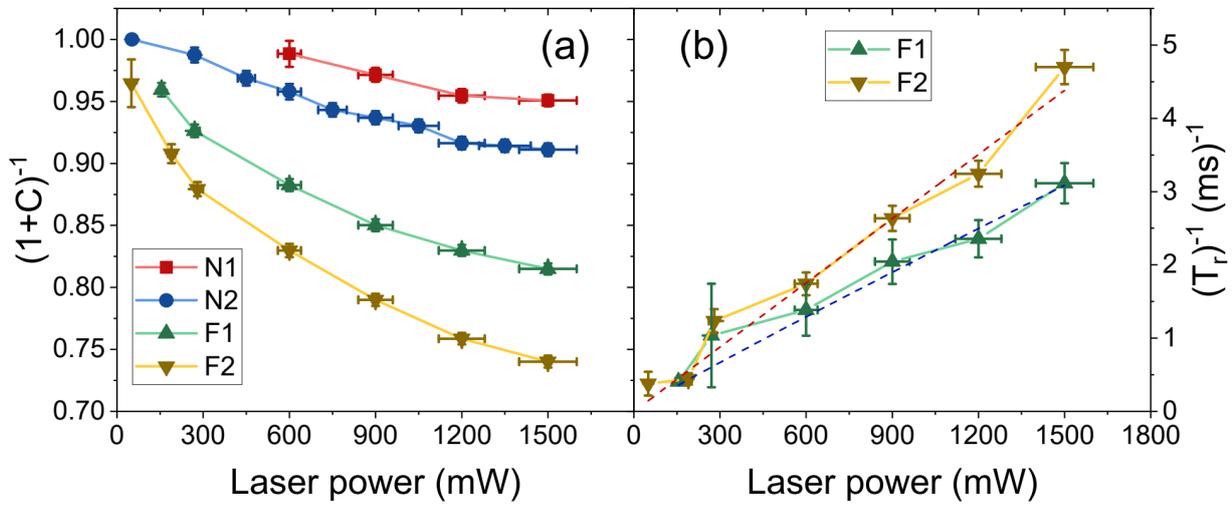

*Figure 4 – Parameters describing $NV^0 \rightarrow NV^-$ photoconversion under laser irradiation. The quantity $(1+C)^{-1}$ in (a) indicates the ratio between the equilibrium value of FL and the initial value, right after the initialization pulse. The decrement is more consistent for the flat samples (down to 75% for sample F2). The photoconversion rate $(T_r)^{-1}$ displayed in (b) is nearly linear with laser power, following a power law with exponents of 0.95±0.07 for F1 and 1±0.07 for F2 (red and blue dashed curves).*

To illustrate the symmetry between the dynamics of the two charge states in the dark, in Fig.5a we show the FL signal in the 550-600nm and 750nm+ spectral windows, dominated by $NV^0$ and $NV^-$, respectively, for sample F2 at the highest laser power. The experiment of Fig.3c was performed with or without an applied magnetic field of 750 G (open and solid symbols in Fig.5a, respectively). Indeed, while the evolution of the $NV^0$ signal reflects strictly charge dynamics, the $NV^-$ signal is influenced by both charge and spin dynamics. The applied magnetic field suppresses spin dynamics, and makes it possible to selectively measure the effects of charge switching for $NV^-$. Under these experimental conditions, the two curves, in red and blue, representing charge state recovery for $NV^-$ and $NV^0$, appear symmetrical, thus corroborating the idea that laser light converts one state into the other.

Performing the experiment of Fig.3c with the two 550-600nm and 750nm+ spectral windows makes it possible to estimate the charge-state ratio R=[$NV^-$]/[$NV^0$] at steady state, as proposed by Giri et al.[34] (see Appendix III). The two assumptions are (i) all the NV centers are optically active and (ii) the sum of $NV^-$ and $NV^0$ is always constant. The latter hypothesis implies that a reduction of $NV^0$ is counterbalanced by an increase of the $NV^-$, and vice-versa. Figures 5b,c show the subdivision of the populations between $NV^-$ and $NV^0$ after a 10 ms laser pulse at low and high power for sample F2, as an example. In both cases, the initialization pulse produces an excess of $NV^-$ centers that convert back into $NV^0$ during the dark time. At low laser power (50 mW, ≈5 mW/μm$^2$), the signal is dominated by $NV^0$s, with a ratio R of 0.13/0.87≈0.15 at the longest timepoint acquired (100 ms). However, at high power (1500 mW, corresponding to ≈150 mW/μm$^2$) the ratio changes to 0.53/0.47≈1.13, with the $NV^-$ centers now being the dominant state after a seven-fold increase compared to the low power case. Interestingly, after an initial rapid decrease, a large value of R (1.13) is observed after 100 ms, and the curve exhibits a much more slowly decaying component. Thus, the effect of an intense laser pulse is to create a large $NV^-/NV^0$ ratio that is sustained over tens or hundreds of milliseconds (see Discussion). Evidently, the dynamic of charges has fast and slow components, summarized by the stretched exponential behavior, with the fast components in the 10-1000 μs timescale, and the slower ones on a 100 ms scale (Fig.5b,c). On the contrary, the timescale of spin dynamics is of the order of ≈1 ms. For this reason, when considering the combined effect of spin and charge dynamics, there will be a substantial overlap in the fluorescence up to 1-10 ms, while at longer times the FL profile will be dominated by charge dynamic. This is also apparent from Fig.5a, where the difference between the two red curves, attributed to spin dynamics, vanishes after ≈10 ms; the evolution at longer times is unaffected by the magnetic field, thus indicating that it can be attributed exclusively to charge dynamics.

Finally, repeating the same experiments with sample N2 gives similar results, but with a modest degree of $NV^0 \rightarrow NV^-$ photoconversion. Even at the highest power of 1500 mW, the ratio $NV^-/NV^0$ is 0.12/0.88≈0.14, much lower compared to the results of sample F2.

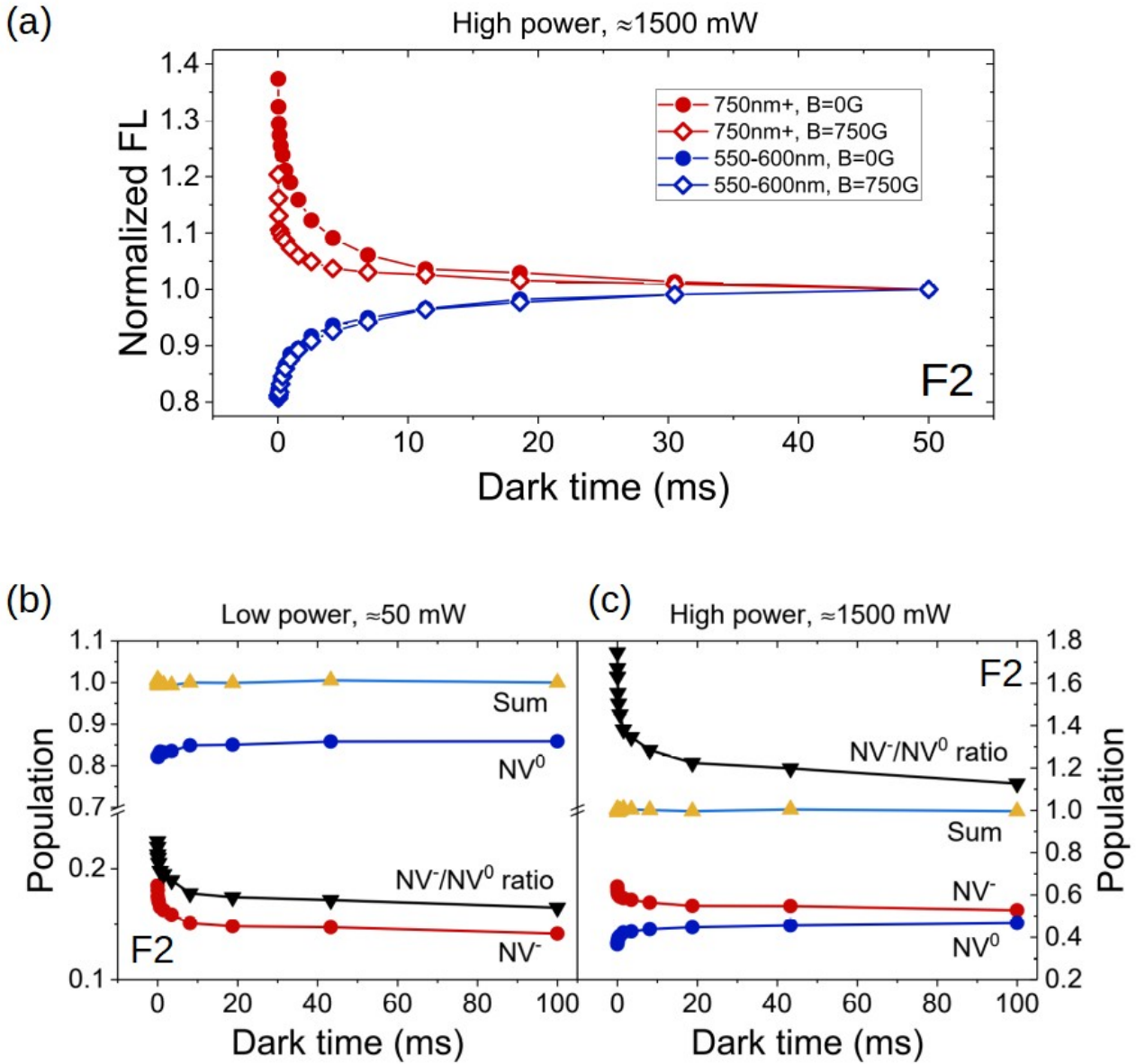

*Figure 5 – **Spin and charge dynamics after the laser pulse.** The effect of a strong magnetic field (750 G) on the fluorescence of sample F2 at a laser power of 1500 mW is shown in (a). FL was collected in the 550-600nm region (blue curves) and in the 750nm+ window (red curves), with and without magnetic field (empty and solid symbols, respectively). When considering the $NV^0$ centers only (550-600nm window), no change in the FL profile is observed, irrespectively of the magnetic field applied. On the contrary, the signal in the 750nm+ window is reduced by a strong magnetic field. This difference is the result of quenching $NV^-$ spin polarization through spin-states-mixing. However, beyond ≈10 ms, the difference vanishes, and the evolution at longer times is dominated by charge dynamics. At low (b) and high (c) laser powers, the 10 ms laser pulse creates an excess of $NV^-$s (red curve) that convert to $NV^0$s (blue curve) in the dark. $NV^-$ and $NV^0$ populations are normalized such that their sum is 1 (yellow-azure curve). The ratio R is indicated by the black curve. A near-equilibrium ratio (0.15) is reached after 100 ms at low power (50 mW), while a large and sustained R (1.13) value is observed after high power irradiation (1500 mW).*

Finally, we report optically detected magnetic resonance (ODMR) experiments to assess the spin polarization of the ground-state level $m_s=0$ of NV$^-$. Continuous laser irradiation polarizes the $m_s=0$ state, while a sweeping microwave field transfers the population to the $m_s=\pm 1$ states. However, we note that the detected ODMR contrast is also affected by NV$^0$ centers. Indeed, even though FL is collected in the 750nm+ window, where most of the signal comes from the NV$^-$ centers, a residual background signal from NV$^0$s is still present. This residual NV$^0$ signal is not modulated by microwave irradiation and cannot be removed entirely by spectral filtering. Hence, we expect the ODMR contrast to be increased at higher laser power due to two independent mechanisms: (i) increased NV$^-$ polarization; (ii) reduction in the nuisance background signal from NV$^0$.

In Fig.6a we show the ODMR spectra from sample F2 for a fixed microwave power and increasing laser power up to 1500 mW. The contrast (i.e. the depth of the dip) increases rapidly with laser power, and makes it possible to resolve the strain-split doublet[40] and the $^{13}$C sidebands at the highest contrast-to-noise ratio. In Fig.6b we show the dependence of the ODMR contrast on laser power for all samples. Nanostructured samples exhibit low contrast and limited gain with increasing laser power, consistent with the evidence that NV$^0$ remains the dominant charge state even at the highest laser powers. The ODMR contrast in flat samples is much larger and increases sharply with laser power. Saturation of contrast for the F2 sample at around 600 mW probably reflects laser-induced repolarization related to insufficient microwave power (see Discussion).

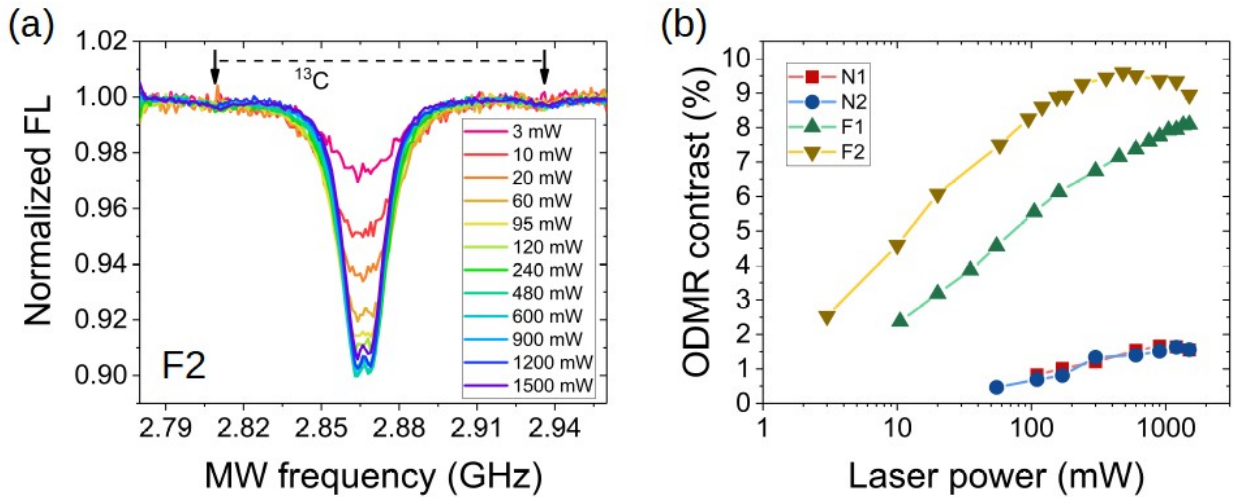

*Figure 6 – **Optically Detected Magnetic Resonance.** (a) ODMR spectra from sample F2 as a function of laser power. (b) ODMR contrast for all samples. The increase in contrast is more pronounced for flat samples.*

**Discussion**

Our results highlight the importance of $NV^0$ to $NV^-$ charge conversion in shallow defects, and point to the use of high-power laser irradiation to improve the availability and polarization of shallow $NV^-$. This effect appeared to be more prominent in flat samples, compared to nanostructured samples.

In all four samples investigated, the $NV^0$ centers represent a substantial proportion of NV defects. In the bulk, the presence of electron donors, such as substitutional nitrogen, favors formation and stability of $NV^-$ center. Near the surface, acceptor states decrease the stability of the $NV^-$, and $NV^0$s may become the dominant charge state[41,42], with detrimental effects for sensing and polarization transfer applications. Our samples were engineered to present shallow NV centers, and it is not surprising to find predominantly $NV^0$ centers. In the case of nanostructured samples, the surface effects are exacerbated by the large surface-to-volume ratio of the nanostructures. Therefore, diamonds with nanopillars have fewer $NV^-$s compared to $NV^0$s, as apparent in the Raman, FL and ODMR spectra (Figures 2c, 4a and 6b, respectively).

Moreover, the integrated fluorescence of Figure 2d suggests that nanostructured samples have a lower amount of NV centers altogether, independent of the specific charge state. This could be determined by less efficient ion implantation on the nanostructures or to a loss of nitrogen due to ion straggling.

Laser irradiation has two effects on the NV centers. The first effect is polarization of the $NV^-$ $m_s=0$ spin state. The second effect is to convert part of the $NV^0$s into $NV^-$ and increase the ratio of the negatively charged centers over the total NVs. Upon switching the laser off, a process of recharging in the dark results in a $NV^- \rightarrow NV^0$ back-conversion. An interesting finding of this work is that the excess $NV^-$ population is persistent over a relatively long timescale.

The following hypothesis can qualitatively explain the experimental findings. A laser pulse ionizes the substitutional nitrogen atoms[43], and electrons are released into the conduction band and recombine with electron-trapping defects[44], including the $NV^0$. The decrease of $NV^0$ FL in Figure 3b is consistent with this mechanism. The phenomenon of redistribution of charges was recently investigated from a different perspective, namely the Stark shift of optical transition[37]. The nearly-linear dependence of $(T_r)^{-1}$ on laser power (Fig.4b) suggests a single-photon ionization process of substitutional nitrogen as a source of electrons for $NV^0 \rightarrow NV^-$ photoconversion. We note that the proposed mechanism for photoconversion is different from the two-photon excitation of electrons from valence band[36], which is better described by a quadratic dependence of $(T_r)^{-1}$ on laser power[20]. In the dark, a $NV^- \rightarrow NV^0$ conversion is observed. The mechanism proposed for this process is electron tunneling which critically depends on the NV environment. Tunneling to a neighboring trap can only happen if the trap is empty[24]. At weak laser powers, the photoexcited electrons do not saturate the traps in the vicinity of the NV center. At stronger laser powers, increasingly distant traps are populated. Hence, the time the system takes to return to equilibrium depends on laser power, as tunneling to distant traps is a much slower process. This picture is consistent with the results of Fig.5b and 5c that show, at high laser power, a large $[NV^-]/[NV^0]$ ratio persistent on a scale of one hundred milliseconds. Thus, high-power irradiation increases the availability of $NV^-$ and extends the time window in which the NV⁻ spin states can be manipulated and detected, with important implications for sensing and polarization-transfer applications.

The beneficial effects of high-power irradiation are less pronounced in the etched samples, with a modest increase in NV⁻ population. This is consistent with the idea that surface defects compete with NV⁰ to capture released electrons. Therefore, the NV⁰→NV⁻ conversion is more efficient in the flat samples and less efficient in the nanostructured ones, where the amount of surface defects is large compared to the amount of NVs (Figure 4a). In our samples, the advantage of increased exposed area resulting from nanostructuring appears to be mitigated by the less favorable charge dynamics. Additional treatment of the surface to stabilize NV⁻ may improve the performance of nanostructured samples, but we have not explored effects of different terminations in this paper.

We used stretched exponentials to fit all the fluorescence curves to account for heterogeneity in the ensemble of NVs, resulting in a superposition of many single exponential curves[8,23,24,45]. The dynamics of charge tunneling was addressed by Miller[46] and Tachiya[47] by considering a spatially homogeneous distribution of charge acceptors and recently extended to a discrete distribution[22]. In this model, the charge dynamics can be described by a functional dependence of the form

$$I(t) = I_{eq}\left(1 + C\exp\left\{\frac{-4\pi a^3 n}{3}[\ln(vt)]^3\right\}\right) \qquad \text{Eq. 2}$$

Here $n$ is the density of trapping defects, and $v$ and $a$ are parameters linked to the NV⁻ center, defined as the tunneling frequency and the attenuation length of the wavefunction, respectively. Using this expression for the fits, we find $a$=0.1 nm, $v$=(0.8±0.1) MHz, $n$=2-4 ppm. The density $n$ shows a slight tendency to decrease after long laser pulses, possibly indicating that photoexcited electrons have filled some charge traps. Additional work will be necessary to understand the features of $n$ and some subtleties not captured by the present model, such as non-uniform values of $n$, $v$ and $a$ across a defective sample. However, for our data, we do not observe a substantial improvement of the fit using Eq.2 compared with a stretched exponential function (Eq.1). We are mostly interested in the initial out-of-equilibrium values of the curves compared to equilibrium. These values are set during the initialization phase by the laser pulse length and power, and do not depend critically on the choice of fitting function. In this respect, the values returned by the two fitting procedures are in good agreement.

The leakage of the NV⁰ FL component in the NV⁻ spectral region can affect the ODMR contrast[48–50]. For NV⁻ centers, the contrast is defined as $C_s = (I_{off} - I_{on})/I_{off}$, where $I_{off}$ and $I_{on}$ are the collected FL intensities with microwaves off- and on-resonance, respectively. If the intensity contains an NV⁰-related component $I_0$, then the contrast $C_s$ must be scaled by a factor $I_{off}/(I_{off}+I_0)$. In our case, the intensity $I_{off}$ ($I_0$) is the fraction $f^-$ ($f^0$) of the total NV⁻ (NV⁰) intensity filtered by the 750nm+ longpass and normalized by photon-counter efficiency. It is possible to relate these intensities to the charge-state ratio R through parameters pertaining to NV⁻ (NV⁰) centers, such as the absorption cross sections $\sigma^-$ ($\sigma^0$), and the excited state lifetime $\tau^-$ ($\tau^0$). The ODMR contrast for the NV ensemble is

$$C_{ens} = C_s \frac{R}{R+F} \qquad \text{Eq. 3}$$

where $F = \frac{f^0}{f^-} \frac{\sigma^0}{\sigma^-} \frac{\tau^-}{\tau^0}$ is a numerical factor that depends on detection conditions, such as the choice of filters and detectors, and structural properties of NV centers. Conversely, R can vary for different samples. For shallow NVs, where the charge state is dominated by neutral centers, R is small and the ensemble contrast is substantially reduced. On the contrary, for samples that mostly contain NV$^-$ centers (as usually happens for bulk NV centers) R is much larger than F and the correction is minimal. For our setup we estimate $f^0/f^- \approx 0.42$ from the shape of fluorescence (see Appendix I). Using $\sigma^- = 3.1 \times 10^{-17}$ cm$^2$ [51], $\sigma^0 = 1.8 \times 10^{-17}$ cm$^2$ [52], $\tau^- = 12$ ns, $\tau^0 = 21$ ns [53], we find F≈0.14. Also, the charge state ratio varies dramatically from 0.15 at low laser power to 1.13 at high power. With an indicative value of $C_s = 10\%$, we estimate ensemble contrasts of 5.2% and 8.9% at low and high laser power, respectively, and a tendency to increase with R and with laser power, following the NV$^0 \to$ NV$^-$ photoconversion. In principle, $C_s$ can be recovered *a posteriori* from experimental values of the ODMR contrast, provided that the dependence of R on the laser power is known. The proposed interpretation explains the trend shown in Fig.6b from low to moderately high laser powers, although the contrast shows a saturation or even a slight reduction at the highest power levels (see for instance sample F2). It also explains the striking difference between the flat and the nanostructured samples: in the latter, NV$^0$s are by far the dominating species even at high laser power and the extracted contrast is always poorer than in the flat samples. Concerning the saturation of the contrast at high powers, we note that an optimal contrast results from combined laser and microwave spin initialization. If the laser power is too high, the microwave field cannot compete in inverting the population levels in the ground state, and some degree of laser-induced repolarization is expected[54]. Despite setting the microwaves power to the maximum available value (output of 8 dBm amplified by ≈35 dB), the laser repolarization effect may prevail, resulting in the saturation and slight reduction shown in Fig.6b.

The resulting increase in ODMR contrast improves sensitivity η (expressed in $T/\sqrt{Hz}$) and the minimum magnetic field δB detectable in a time Δt, defined as $\delta B = \eta/\sqrt{\Delta t}$. In continuous-wave ODMR protocols, sensitivity is given by the relation[55]

$$\eta = \frac{4}{3\sqrt{3}} \frac{h}{g\mu_B} \frac{\Gamma}{C_{ens}\sqrt{r}} \qquad \text{Eq. 4}$$

where the numerical factor in front is related to the Lorentzian function used to fit the resonance, $h$ is the Planck constant, $\mu_B$ is the Bohr magneton, $g \approx 2$, $\Gamma$ is the full-width-half-maximum of the Lorentzian, and $r$ is the photodetection rate. The contrast $C_{ens}$ in Eq.4 is the ODMR contrast of the NV ensemble given by Eq.3. At low laser power (50 mW), we find experimentally r = 0.73x10$^6$ s$^{-1}$, while at high power (1500 mW) we find r = 17.8x10$^6$ s$^{-1}$ (values normalized by limited photon counter efficiency of the detector). Taking values of $C_{ens}$ and R as before, and a representative $\Gamma$ = 10 MHz, we find $6.2\,\mu T/\sqrt{Hz}$ and $730\,nT/\sqrt{Hz}$ at low and high power, respectively. Sensitivity might be further improved with the use of pulsed-wave ODMR[55] and extended to the detection of AC magnetic fields[56,57]. Therefore, magnetometry applications benefit from the use of strong laser power through an increase in photon detection rate and improved contrast. We shall note that the correction factor of Eq.3 does not exclusively determine ODMR contrast, but plays a role in other detection schemes, including pulsed sequences as those in Fig.3a and 3c.

$^{13}$C nuclear hyperpolarization is another technique that could benefit from an increased fraction of NV⁻s. This technique is based on the transfer of spin polarization from highly-polarized NV⁻ to first-neighbor $^{13}$C nuclei via spin-spin interaction at specific values of the applied magnetic field[58]. Polarization can then diffuse through the diamond lattice to the bulk $^{13}$C spin reservoir by nuclear spin-spin interactions[59,60]. However, this process is much slower compared to the fast hyperpolarization of the first shell $^{13}$C nuclei (≈1 ms) in the vicinity of NV⁻ centers, and increasing the amount of NV⁻s via photoconversion can substantially improve the efficiency of the hyperpolarization process.

The effects reported here may be particularly important for hyperpolarization of nuclei outside the diamond surface. Indeed, it has been proposed that shallow NV⁻ may provide a source of spin polarization of molecular moieties adsorbed at the diamond surface[16,61], thus complementing techniques like Dynamic Nuclear Polarization or optical pumping spin-exchange. Increasing their concentration in the vicinity the diamond surface may enable hyperpolarization of nuclei outside the diamond lattice.

In conclusion, we studied the effects of laser power on charge-state photoconversion in ensembles of shallow NV centers in flat and nanostructured diamond. Surface effects favor the neutral charge state (NV⁰) over the negative state (NV⁻). This problem is exacerbated in the nanostructured samples. We show that high-power laser irradiation promotes NV⁰ to NV⁻ photoconversion, thus significantly increasing the relative abundance of NV⁻s in flat and in nanostructured diamond. Interestingly, we find that the excess NV⁻ population is stable over time scales of tens or hundreds of milliseconds. This may provide an important advantage for magnetometry and polarization-transfer applications, where large ensembles of shallow NV⁻ are needed.

**Appendix**

*I – Filtering of fluorescence signal*

Fluorescence is collected with a combination of bandpass filters, the 550-600nm and the 750nm+ (purple and blue curves in Figure 7, respectively), with peak transmittance of 80-90%, indicatively. The emission spectrum of the nanostructured sample N1 (black curve) was acquired with a confocal Raman setup at low power (0.41 mW/µm$^2$). The fluorescence spectrum can be decomposed in the NV⁰ and the NV⁻ bands (green and red curves): the 550-600nm filter selects almost exclusively the NV⁰ signal, but the 750nm+ filter selects both signals. Thus, the emission from NV⁰ centers cannot be completely removed by spectral filtering and it can affect the NV⁻ signal by providing an undesirable background signal (compare Eq.3). However, at high laser power the NV⁻/NV⁰ ratio increases (Fig.2) and the leakage becomes less relevant.

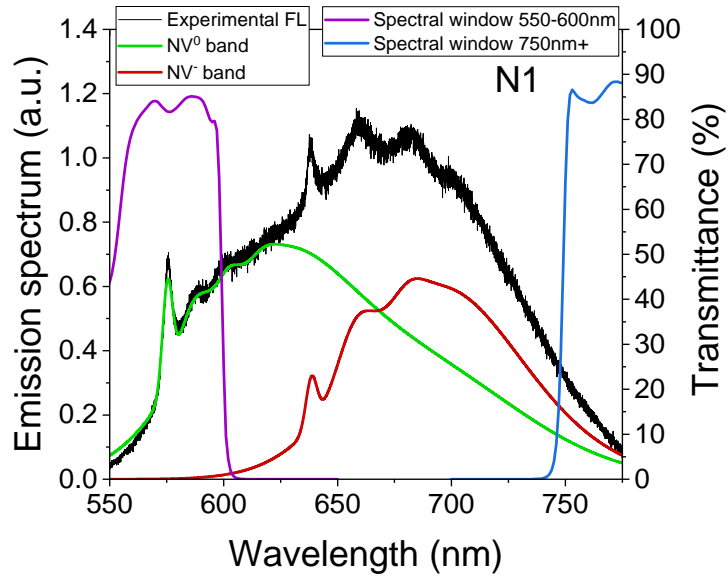

*Figure 7 – Filtering of fluorescence. The emission spectrum of N1 (black curve) is collected with the Raman confocal microscope and can be decomposed into $NV^0$ and $NV^-$ components (green and red curves, respectively). Filter 550-600nm selects only the $NV^0$ component, but both components pass through the filter 750nm+. The leakage of $NV^0$ FL in the $NV^-$ region decreases the contrast in ODMR and pulsed laser experiments. It is preferable to use a high laser power, as it increases the amount of $NV^-$s by photoconverting $NV^0$s and reduces the $NV^0$ leakage.*

*II – Effect of the readout pulse*

In most of the experiments, a readout pulse of 5 μs was adopted. We considered the possibility that even a 5 μs readout pulse may induce photoconversion, especially at high laser power. In Figure 8 we plot the evolution of the FL signal in the dark, describing the spin and charge dynamics of sample F1 (i.e., no magnetic field applied, 750nm+ filter, analogous to the red solid curve of Fig.5a) at the highest laser power of 1.5 W, with "short" and "long" readout pulses of 0.5 μs (green curve) and 5 μs (purple curve), respectively. We did not observe substantial difference between the two curves. However, the acquisition time was affected by the length of the readout pulse (since the photon-shot noise for a Poissonian process scales as the inverse square root of the number of photons detected, or, equivalently, as the inverse square root of the acquisition time). By way of example, the green curve was acquired after 220 repetitions, while the purple curve only needed 50 repetitions. We conclude that a 5 μs readout pulse ensures a good sensitivity without substantially perturbing the distribution of NV charge states even at the highest laser power.

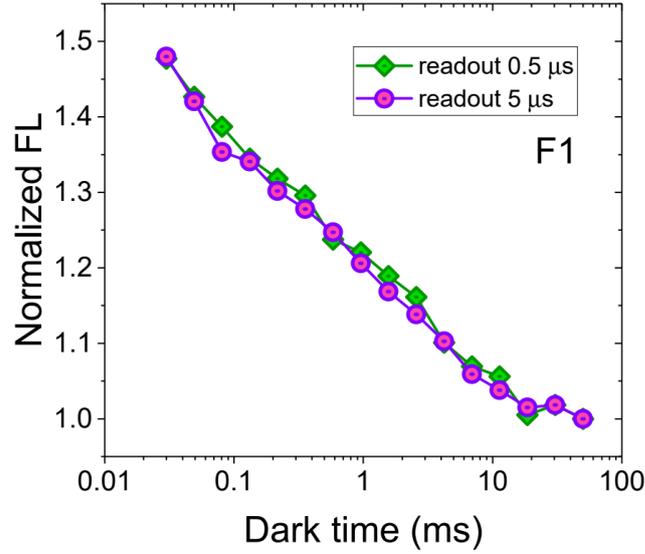

*Figure 8 – **Effect of the readout pulse.** Two readout pulses of 0.5 µs (green curve) and 5 µs (purple curve) were used to record the FL variation in the dark in a pulsed experiment where both charge and spin dynamics were investigated (no magnetic field applied, 750nm+ filter). Laser power was 1.5 W, the maximum available. Even under these conditions the curves do not display any major difference. Results are from sample F1.*

### III – Calculation of the charged state ratio

An important parameter that characterizes the samples is the ratio between the concentrations of negatively centers to neutral centers, defined as R=[NV⁻]/[NV⁰]. A simple pulse sequence (Fig.3c in the main text) makes it possible to estimate this ratio[34]. To this end, a strong magnetic field (750 G) is applied along the [100] direction to suppress the spin polarization mechanism, and selectively investigate charge dynamics. Indeed, a sufficiently strong magnetic field (>600 G) mixes the spin states and quenches the mechanism responsible for the $m_s$=0 polarization, leading to a reduction of the spin-related component in the fluorescence. We then record fluorescence through two spectral windows, the 550-600nm (for the NV⁰s) and the 750nm+ (for the NV⁻s). The two signals are fitted by stretched exponentials.

Only NV⁰ signal passes through the 550-600nm filter and the FL increases with time as

$$I^0(t) = I^0_{eq}\left[1 - \delta_0 e^{-(t/T_r)^n}\right] \qquad \text{Eq. A1}$$

On the contrary, a fraction of the NV⁰ fluorescence leaks through the 750nm+ window. The composite signal decreases with time as:

$$I(t) = k I^0_{eq}\left[1 - \delta_0 e^{-(t/T_r)^n}\right] + I^-_{eq}\left[1 + \delta_- e^{-(t/T_r)^n}\right] \qquad \text{Eq. A2}$$

Here, $I_{eq}^-$ and $I_{eq}^0$ are the levels of fluorescence at equilibrium, $T_r$ and n govern the dynamics of charge conversion, and $\delta_0$, $\delta_-$, are positive parameters that indicate the initial out-of-equilibrium level of fluorescence. The number k quantifies the amount of $NV^0$ leakage, and it can be estimated from the FL profile and the spectral windows transmittivity (here k≈0.19). From Eq.A1 and knowing k, it is possible to recover the pure $NV^-$ signal in Eq.A2.

Reasonably, the total number of NV centers, either in the neutral or in the negative charged state, is not affected by the laser pulses (*i.e.* we do not consider the potential contribution of $NV^+$ [62], which are only present in very low concentration[63]). Thus, a reduction in $N^-(t)$, the number of $NV^-$s, is counterbalanced by a gain in $N^0(t)$, the number of $NV^0$s, such that $N^-(t)+N^0(t)=N_{tot}=cnst$. With this assumption, the fractions of $NV^-$ and $NV^0$ centers over the total at equilibrium are $\frac{N^-_{eq}}{N_{tot}}=\frac{\delta_0}{\delta_-+\delta_0}$ and $\frac{N^0_{eq}}{N_{tot}}=\frac{\delta_n}{\delta_n+\delta_0}$. The charge state ratio is then given by $R=\frac{N^-_{eq}}{N^0_{eq}}=\frac{\delta_0}{\delta_-}$.

## Acknowledgments

We thank P. Olivero for useful suggestions on sample treatments in the department of physics at University of Torino, Italy. We also thank L. Basso for Raman spectroscopy in the department of physics at Università di Trento, Italy. Finally, we thank Simone Lauciello and Alice Scarpellini of Electron Microscopy Facility of Istituto Italiano di Tecnologia in Genova for SEM characterization.

This project has received funding from the European Union's Horizon 2020 research and innovation programme under grant agreement No 858149 (AlternativesToGd), and from the Marie Skłodowska-Curie program, Grant Agreement No. 766402 (ZULF).